\documentclass[twocolumn,raggedbottom]{aastex701}
\usepackage{graphicx}
\usepackage{amsmath}
\usepackage{amssymb}

\shorttitle{Double Transients in Galactic Nuclei}
\shortauthors{Kritos \& Perna}

\begin{document}

\title{Tidally Induced Coda: Double Transients from Black Hole--Star\\ Binary Encounters with Massive Black Holes}
\correspondingauthor{Konstantinos Kritos}
\email{konstantinos.kritos@stonybrook.edu}

\author[0000-0002-0212-3472]{Konstantinos Kritos}
\altaffiliation{NASA Hubble Fellow}
\affil{Department of Physics and Astronomy, Stony Brook University, Stony Brook, NY 11794-3800, USA}
\email{konstantinos.kritos@stonybrook.edu}
\author[0000-0002-3635-5677]{Rosalba Perna}
\affil{Department of Physics and Astronomy, Stony Brook University, Stony Brook, NY 11794-3800, USA}
\email{rosalba.perna@stonybrook.edu}

\begin{abstract}

Dense stellar environments surrounding massive black holes (MBHs) provide a natural setting for dynamical interactions that can produce transient electromagnetic and gravitational-wave sources. In particular, the tidal field of an MBH can strongly perturb binaries undergoing close encounters with it, driving their internal orbits to extreme eccentricities. Here, we investigate the outcomes of such interactions for binaries composed of a stellar-mass black hole (BH) and a star. Using direct few-body scattering experiments with the code \href{https://github.com/YihanWangAstro/SpaceHub}{\tt SpaceHub}, we explore hyperbolic encounters between BH--star binaries and MBHs. We systematically quantify the probabilities of different dynamical outcomes as functions of the binary orbital parameters and MBH mass, with particular emphasis on a channel in which the tidally induced eccentricity triggers a micro-tidal disruption event ($\mu$TDE), as the star is disrupted by its BH companion, followed by accretion of the stellar debris onto the MBH. We derive fitting formulae for the $\mu$TDE cross section and for the time delay between the two resulting flares, finding that the latter can be a few months for an MBH of $10^6\,M_\odot$. These double transients offer a distinctive observational signature of $\mu$TDEs. We estimate event rates ranging from a few to $\sim 100\,{\rm yr}^{-1}$ within a luminosity distance of $40\,{\rm Gpc}$.

\end{abstract}


\section{Introduction} 
\label{sec:intro}

Galactic nuclei host massive black holes (MBHs) whose gravity dominates the dynamics of stars within a characteristic sphere of influence. Two-body relaxation continuously scatters stars onto low-angular-momentum ``loss-cone'' orbits, a fraction of which bring a star close enough to the MBH to be tidally disrupted, producing the luminous flares known as tidal
disruption events~\citep[TDEs;][]{Rees1988}.~Wide-field
time-domain surveys have now identified over a hundred TDE candidates~\citep{Gezari:2021bmb,Langis2026}, with facilities such as the Vera C.\ Rubin Observatory expected to increase this sample by orders of magnitude~\citep{Karmen:2026ixp}.

A substantial fraction of stars in nuclear star clusters (NSCs), however, are not single but reside in binaries.~When a binary plunges onto a nearly radial orbit toward the MBH, the tidal field acts simultaneously on the binary orbit and on the individual binary members, producing a much richer set of outcomes than a single-star TDE: the MBH tide can unbind the binary, ejecting one member as a hypervelocity star while capturing its companion onto a tight orbit around the MBH~\citep{Hills1988}; it can
tidally disrupt one or both binary members; or, for sufficiently tight binaries, the same tidal torque that would otherwise separate the pair can instead drive the two stars into a direct physical collision near pericenter passage~\citep{Ginsburg2007,Antonini2010,Antonini2011,Bradnick2017}. \citet{YuLai2024} used three-body scattering experiments to map these outcomes as a function of  the ratio $\beta$ of the binary's tidal radius to its pericenter distance from the MBH, finding that collisions occur with probabilities of a few to tens of percent depending on $\beta$.

We note that an additional channel operates when binaries are already bound in an elliptical orbit to the MBH. In this case, the MBH and the binary form a hierarchical triple, and the tidal torque exerted by the MBH can drive von Zeipel--Lidov--Kozai oscillations of the binary's eccentricity and inclination~\citep{Kozai1962,Lidov1962,2019MEEP....7....1I}. For sufficiently favorable orbital configurations, the binary can reach extreme eccentricities, at which point processes that are negligible at moderate eccentricity, such as stellar collisions or gravitational-wave emission, can become dominant.
Previous studies have shown that this mechanism can induce mergers of stellar binaries in the Galactic Center and of compact-object binaries embedded in NSCs~\citep{AntoniniPerets2012,Fragione2019, Fragione2020,Maeda2026,Gao2026,WangH2026}.

Here we perform the first study (to the best of our knowledge) of the outcomes of BH--star binaries approaching MBHs, for a wide range of MBH masses,  in  a single hyperbolic passage from the nuclear cluster's loss cone. As the eccentricity of the inner binary increases, the pericenter distance between its two components shrinks;  this can result in a partial or full tidal disruption of the star by its BH binary companion, i.e., a micro ($\mu$)TDE~\citep[e.g.,][]{Perets2016}, followed by an `encore,' that is, a secondary flare produced by accretion onto the MBH of the stellar debris that, while unbound from the stellar BH, is still bound to the MBH~\citep{Ryu2024,Johnson2026}. Other potential outcomes include formation of bound triples, direct plunges, ionizations, and exchanges.

Using direct few-body scattering experiments including post-Newtonian corrections, we show that the $\mu$TDEs triggered via these hyperbolic binary encounters lead to a second flare on the timescale of days--months for intermediate-mass BHs, and months--years for low-mass supermassive BHs, and provide fitting formulae for both the $\mu$TDE cross section and the time delay between the two flares. This newly identified class of transients can help  us recognize the electromagnetic signatures of $\mu$TDEs accompanied by better-known classical TDEs, as well as probe the number densities of BH--star binaries in NSCs.

We describe our numerical methods in Sec.~\ref{sec:methods}, present the simulation 
results in Sec.~\ref{sec:results}, and summarize and discuss our findings in Sec.~\ref{sec:discussion}.

\section{Numerical Methods} \label{sec:methods}

We perform our scattering experiments with the high-precision few-body gravity toolkit \href{https://github.com/YihanWangAstro/SpaceHub}{\tt SpaceHub} \citep{Wang2021}, a C++ code built around algorithmic chain regularization, which offers several integration schemes optimized for extreme mass ratios and extreme eccentricities, both of which are generic features of MBH--binary encounters, and which has previously been used for compact-object scattering in the vicinity of an MBH \citep[e.g.,][]{Wang2021b}. \href{https://github.com/YihanWangAstro/SpaceHub}{\tt SpaceHub} includes post-Newtonian corrections to the equations of motion up to 2.5\,PN order (1\,PN and 2\,PN precession terms and the 2.5\,PN gravitational-wave-emission term) for the relevant compact subsystems, which we include throughout.

Each experiment consists of an inner binary, a stellar-mass BH of $M_{\rm BH} = 10\,M_\odot$ bound to a $M_\star = 1\,M_\odot$ main-sequence star ($R_\star = 1\,R_\odot$), of total mass $M_{\rm bin} \equiv M_{\rm BH} + M_\star$, which is scattered by another BH of mass $M_\bullet$ on an initially hyperbolic orbit, with $M_\bullet$ spanning the comparable-mass, intermediate-mass, and supermassive regimes from $10$ to $10^{10}\,M_\odot$.

Holding $M_{\rm BH}$ and $M_\star$ fixed, we perform $1500$ realizations at each point of a four-dimensional grid spanning the perturber mass $M_\bullet$ (19 half-decade points, $10$--$10^{10}\,M_\odot$), the inner-binary semimajor axis $a_{\rm bin}$ (13 log-spaced points, $0.03$--$30\,\mathrm{AU}$), the velocity ratio $\tilde v \equiv v_\infty/v_{\rm orb}$ (7 log-spaced points, $10^{-3}$--$10$, where $v_\infty$ is the asymptotic encounter speed and $v_{\rm orb}$ the binary circular orbital speed at $a_{\rm bin}$, so $\tilde v \ll 1$ is hard/slow and $\tilde v \gtrsim 1$ soft/fast), and the eccentricity $e_{\rm bin}$ (12 nodes from $0$ to $\min[0.99,e_{\max}(a_{\rm bin})]$, set by requiring the inner pericenter to exceed $R_t=R_\star(M_{\rm BH}/M_\star)^{1/3}$, with finer resolution above $0.9$). 

In each realization the orientation of the binary plane relative to the perturber's orbit is randomized and the stellar orbital phase, or true anomaly $\nu$ (the angular position of the reduced mass relative to pericenter), is set so that the star occupies a uniformly random point in time along its orbit: we first randomly draw the mean anomaly ${\cal M}\in[0,2\pi)$, which represents the angular parameter proportional to the time elapsed since pericenter passage; we then numerically solve Kepler's equation, ${\cal M}={\cal E} - e_{\rm bin}\sin{\cal E}$, for the eccentric anomaly ${\cal E}$, and finally, we solve for $\nu$ using the tangent half-angle formula: $\tan(\nu/2)=\sqrt{(1+e_{\rm bin})/(1-e_{\rm bin})}\tan({\cal E}/2)$.
The impact parameter is sampled with gravitational focusing up to a maximum pericenter $R_{p,\max} = \max(5\,R_{t,\rm bin},\,8\,R_g)$, where $R_{t,\rm bin} = a_{\rm bin}(M_\bullet/M_{\rm bin})^{1/3}$ is the binary's tidal radius about the perturber and $R_g = G M_\bullet/c^2$ its gravitational radius.
Through trial and error we determined that non-flyby outcomes already converge for $R_{p,\max}=5R_{t,\rm bin}$; to avoid a direct relativistic plunge in the $M_\bullet>10^8\,M_\odot$ regime, we impose a floor of twice the marginally bound radius ($8 R_g$) on the maximum pericenter.
Each integration starts at $R_{\rm start}=\max(50R_{t,\rm bin}, 10R_p)$, where $R_p$ is the sampled pericenter of each realization; at this distance, the perturber's tidal perturbation on the binary is~$\lesssim10^{-5}$.

We classify each encounter by its dynamical outcome: \emph{flyby} (the inner binary survives and the perturber recedes), \emph{exchange} (the star or the stellar-mass BH is swapped into a new pairing), \emph{ionization} (all three bodies become mutually unbound), \emph{bound triple} (a hierarchical, unresolved bound triple forms), \emph{plunge} (the star or the stellar-mass BH is captured directly by the nonspinning perturber, within its marginally bound radius $4R_g$), \emph{$\mu$TDE} (disruption of the star by its stellar-mass BH companion), and \emph{TDE by the perturber}.

We classify an encounter as a $\mu$TDE when the star passes within its tidal disruption radius around the stellar-mass BH, $R_t\simeq2.15\,R_\odot \simeq 0.01\,\mathrm{AU}$ (the analogous TDE by the perturber replaces $M_{\rm BH}$ with $M_\bullet$).
For a TDE by the perturber, the star is swallowed whole without an observable flare when $R_\star(M_\bullet/M_\star)^{1/3} \le 4R_g$ (assuming a nonspinning perturber).

Binary separation and exchange are identified when the energy of either binary member becomes negative with respect to the perturber. The encounter is terminated once an escaping body has positive energy relative to the inner pair and its distance from the perturber exceeds $2R_{\rm start}$. Encounters in which no body has escaped within $2\times10^4$ orbital periods $P=2\pi\,[R_{\rm start}^3/G(M_\bullet+M_{\rm bin})]^{1/2}$ are recorded as bound triples.

For each grid cell, given the fraction $f_X$ of realizations ending in outcome $X$, the cross section for that outcome is $\Sigma_X = \pi b_{\max}^2 f_X$ (with $68\%$ binomial confidence intervals), where $b_{\max}^2=R_{p,\max}^2 + 2G(M_\bullet+M_{\rm bin})R_{p,\max}/v_\infty^2$. In total the campaign comprises $31{,}122{,}000$ scattering experiments.

Additionally, we perform a suite of star--star--MBH encounter simulations using \href{https://github.com/YihanWangAstro/SpaceHub}{\tt SpaceHub} to reproduce the results of~\citet{YuLai2024} on the fraction of stellar collisions. This numerical investigation serves three purposes: (i) to provide an independent code comparison, (ii) to assess the importance of PN corrections, and (iii) to compare $\mu$TDEs and stellar collisions as two related classes of dynamically induced astrophysical phenomena.
See Appendix~\ref{app:starstar-collisions}.

\section{Results} \label{sec:results}

\subsection{Branching ratio of various outcomes}
\label{sec:outcomes}

\begin{figure*}
    \centering
    \includegraphics[width=\linewidth]{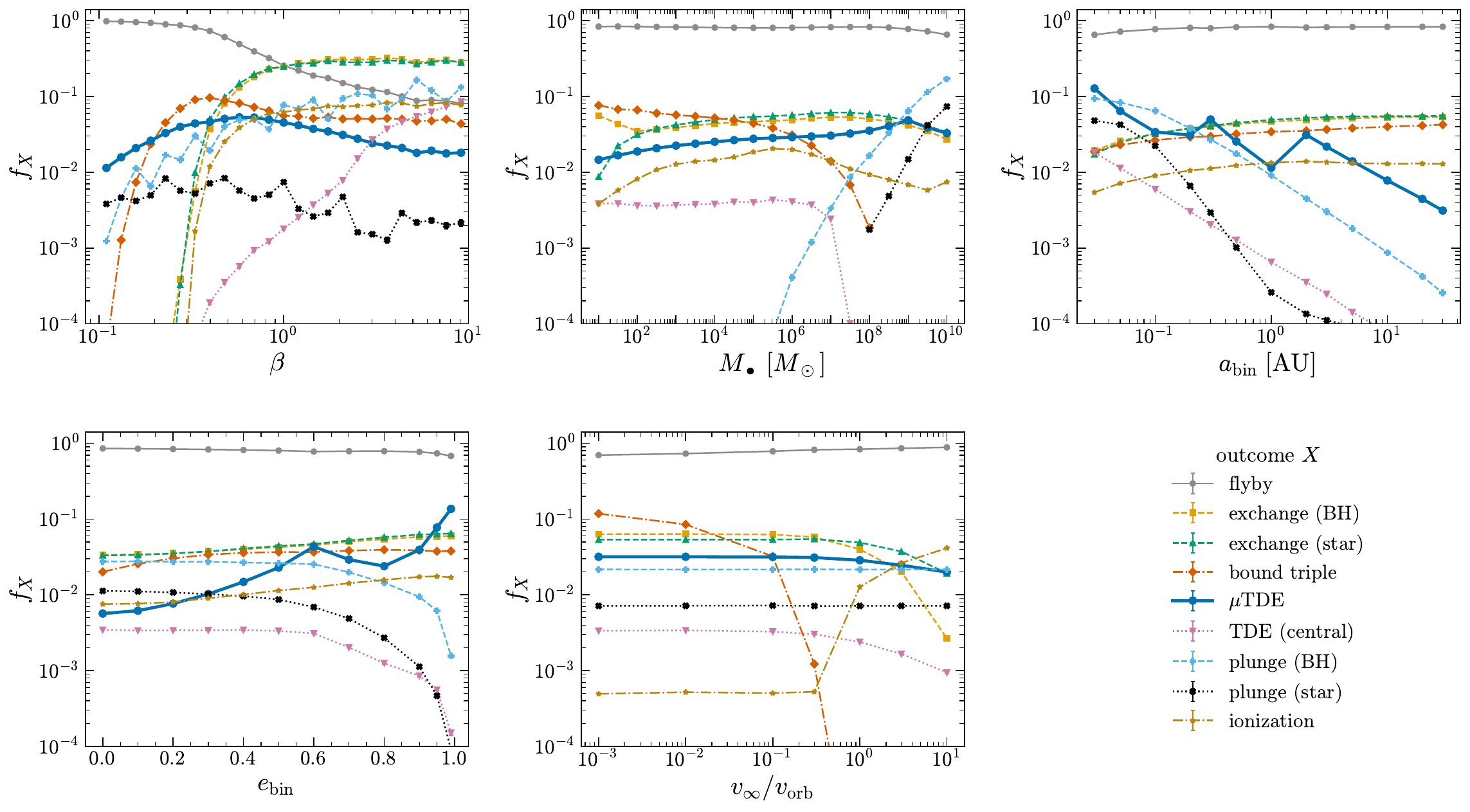}
    \caption{Fraction $f_X$ of outcome $X$ as a function of (from left to right and top to bottom): the penetration parameter $\beta$, the perturber mass $M_\bullet$, the binary semimajor axis $a_{\rm bin}$ and eccentricity $e_{\rm bin}$, and finally the velocity ratio $\tilde{v}=v_\infty/v_{\rm orb}$.  The 68\% binomial uncertainties are too small to see.}
    \label{fig:branching_all}
\end{figure*}

The results of our numerical experiments are summarized in Fig.~\ref{fig:branching_all}, where the fractions $f_X$ of the various outcomes are reported as a function of the binary orbital parameters, perturber mass, and penetration parameter $\beta\equiv R_{t,\rm bin}/R_p$.
The vast majority of our scattering experiments ended with a flyby ($\approx80\%$). About $1\%$ led to complete ionization of the inner binary; $\approx3\%$ ended in a plunge onto the perturber (primarily the stellar-mass BH in the $M_\bullet\gtrsim10^9\,M_\odot$ regime); $\approx3\%$ resulted in the star being tidally disrupted by the stellar-mass BH; $\approx9\%$ resulted in an exchange, with the star captured by the perturber in half the cases and the stellar-mass BH in the other half, producing a hypervelocity object. However, for $M_\bullet\lesssim100\,M_\odot$ most captures involve the stellar-mass BH, whereas for $M_\bullet\gg M_{\rm BH},M_\star$ both members appear light to the perturber and are equally likely to be exchanged.
Finally, in about $3\%$ of all encounters the interaction ended as inconclusive, leading to a hierarchical, unresolved bound triple, which requires longer integration to determine the outcome.

In the highly penetrating regime ($\beta>1$, i.e., $R_p<R_{t,\rm bin}$), there is still a $\sim10\%$ probability that the inner binary survives in a flyby encounter, depending on the binary phase and orientation. However, the most likely outcome in this $\beta>1$ regime is binary disruption, capture by the perturber, and ejection through the Hills mechanism.

The $\mu$TDE outcome fraction becomes important for mildly penetrating encounters and peaks at $\beta\approx0.6$. Furthermore, since $\mu$TDEs occur from eccentricity excitation of the inner binary to the point where its pericenter becomes $<R_t$, the more eccentric the initial binary, the easier it is to excite it to the threshold for a $\mu$TDE outcome. In particular, binaries with an eccentricity arbitrarily close to this critical value require an arbitrarily small perturbation to excite a $\mu$TDE, and thus the corresponding branching fraction $f_{\mu\rm TDE}$ slowly asymptotes to zero as $\beta\to0$. Owing to the high computational cost, we have not simulated highly eccentric binaries ($e_{\rm bin}>0.99$) at small penetration ($\beta<0.2$), so the $\mu$TDE cross section is incomplete there. However, the fraction of binaries with $e_{\rm bin}>0.99$ is only about $2\%$ and $14\%$ under the thermal ($p(e)=2e$) and superthermal ($p(e)=e/\sqrt{1-e^2}$) distributions, respectively, and thus the incomplete region of the parameter space is of small phase volume compared with the bulk where all of our results have converged.

The heavier the perturber, the stronger the tidal influence on the inner binary and the consequent eccentricity increase of the binary. Therefore, our $\mu$TDE branching ratio has also not converged for $M_\bullet\gtrsim10^9\,M_\odot$, for which a $\mu$TDE can be induced also for impact parameters larger than those we sample. However, again, the majority of MBHs have masses $<10^8\,M_\odot$, and thus our rate calculations are not impacted by the incompleteness of the $\mu$TDE cross section for $M_\bullet\gtrsim10^9\,M_\odot$ (where we estimate that our sampling captures $\sim50\%$ of the cross section).

We find that all branching ratios are independent of $v_\infty$ for hard incoming inner binaries (i.e., for $v_{\rm orb}>v_\infty$). For soft binaries, for which $v_{\rm orb}<v_\infty$, ionization becomes significant with a probability of at least $10\%$ and reaching $30\%$ at $v_\infty\approx10\,v_{\rm orb}$. For very hard binaries ($v_\infty\ll v_{\rm orb}$) the ionization fraction drops to a few$\,\times10^{-3}$ and slowly declines with $v_{\rm orb}/v_\infty$. Bound triples occur only for hard binaries and dominate the outcome for $v_\infty/v_{\rm orb}\lesssim10^{-2}$, low $M_\bullet$ ($\lesssim10^4\,M_\odot$), and moderate penetration ($\beta\approx0.4$).

The fraction of TDEs by the perturber drops to zero above $\approx4\times10^7\,M_\odot$, where $R_\star(M_\bullet/M_\star)^{1/3}<4R_g$ and direct plunges dominate. Below this mass, $f_{\rm TDE}\approx4\times10^{-3}$, a factor of $\sim7$ smaller than the $\mu$TDE fraction.

\subsection{$\mu$TDE and debris ballistics}

\begin{figure*}
    \centering
    \begin{interactive}{animation}{figures/mutde.mp4}
        \includegraphics[width=\linewidth]{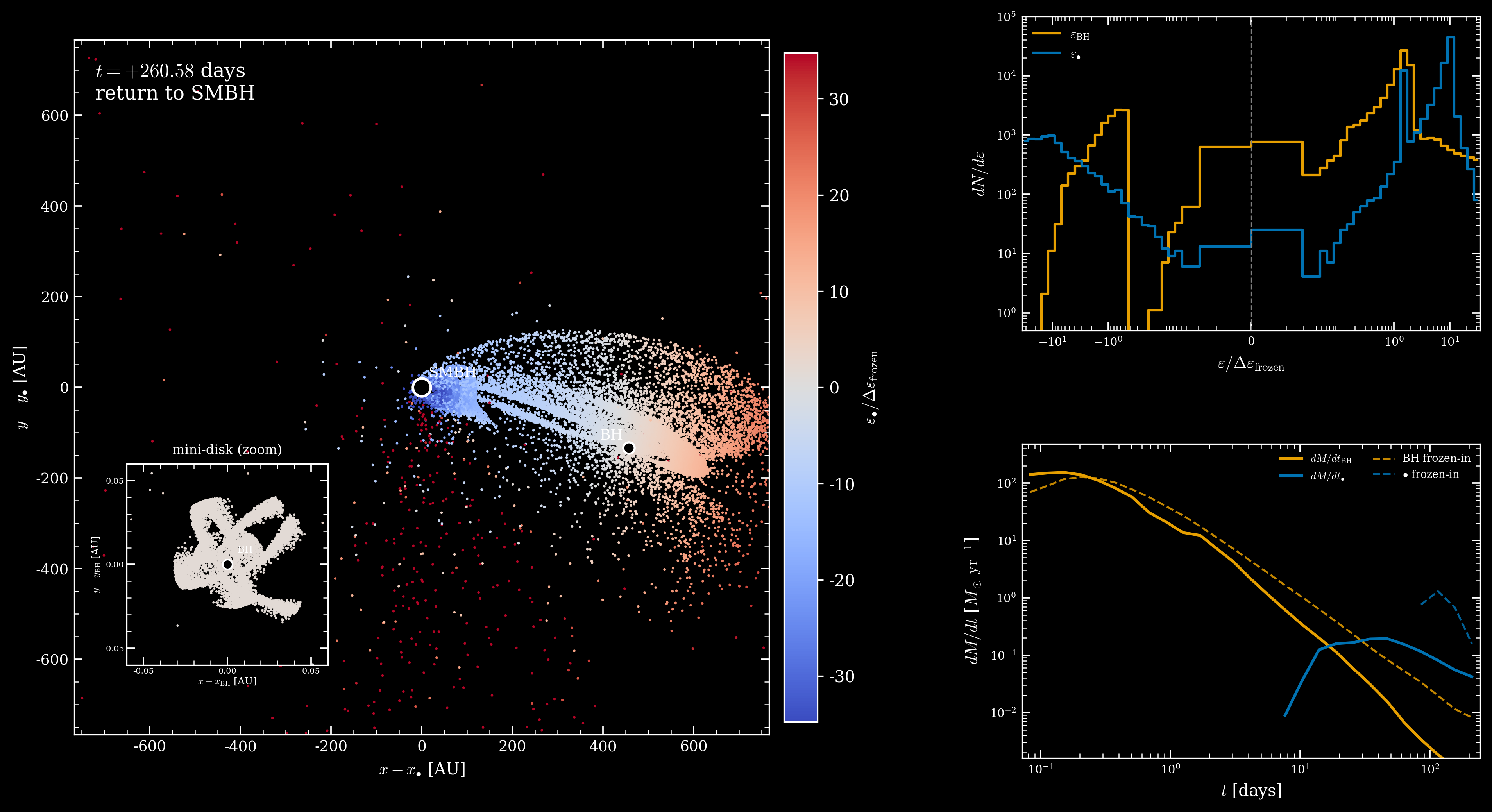}
    \end{interactive}
    \caption{Snapshot of the spatial (left) and specific-energy (top right) distributions of the $10^4$ debris particles at $t=+260.58\,\rm days$ ($\approx8.56\,\rm months$) after the $\mu$TDE. The MBH sits at the center of the left panel (big black dot), and the stellar-mass BH companion (small black dot) is seen receding toward the right after its pericenter passage. The colorbar shows the specific energy of each debris particle normalized to the energy spread at the moment of disruption, $\Delta\varepsilon_{\rm frozen}\equiv GM_{\rm BH}R_\star/R_t^2$: blue for debris bound to the MBH ($\varepsilon_\bullet<0$), and red for particles unbound from the MBH ($\varepsilon_\bullet>0$). The zoomed-in inset in the left panel shows the debris orbital distribution and the mini-disk bound to the escaping BH. Finally, the lower right panel shows the fallback rates onto the BH and the MBH computed with the direct integration method (solid) and the frozen-in approximation (dashed). An animated version of this figure is available in the online article and at~\url{https://youtu.be/g8_UbibAqKQ}. The 19\,s animation follows the debris from the moment of disruption to $t=+260.58\,\rm days$, showing the formation of the mini-disk around the receding BH and the return of the MBH-bound debris.}
    \label{fig:debris}
\end{figure*}

As a representative example of a $\mu$TDE induced via the channel described in Sec.~\ref{sec:outcomes}, we integrate the ballistic trajectories of the stellar debris for $M_\bullet=10^6\,M_\odot$, $a_{\rm bin}=0.1\,\rm AU$, $e_{\rm bin}=0.85$, and $v_\infty=v_{\rm orb}\approx312\,\rm km\,s^{-1}$, corresponding to a nuclear velocity dispersion of $\sigma=v_\infty/\sqrt{3}\approx180\,\rm km\,s^{-1}$.~The orbital angles of this realization are $\iota=3.06878\,\rm rad$, $\Omega=0.445981\,\rm rad$, $\omega=0.347833\,\rm rad$, and $\nu=3.60568\,\rm rad$. We simulate the incoming orbit until stellar disruption with \href{https://github.com/YihanWangAstro/SpaceHub}{\tt SpaceHub}, and use the final conditions to evolve each debris particle forward in time with a Newtonian velocity-Verlet integrator. Specifically, at disruption, we replace the star with $10^4$ test particles and follow their ballistic motion in the time-dependent potential of the two BHs, classifying each as bound to the BH, bound to the MBH, or unbound. Following~\citet{1989ApJ...346L..13E}, each fluid element keeps the position and velocity it had at the instant of disruption, so its depth in the BH's potential sets its specific orbital energy. We sample each particle's position inside the star from an $n=3$ polytrope and assign it the bulk stellar velocity plus a solid-body spin  using the star's angular velocity with respect to its BH companion at
disruption.

We compute the mass fallback rate $dM/dt$ directly: each bound debris particle carries an equal mass $\Delta m=M_\star/10^4$, and its first-return time is recorded when it reaches its point of closest approach to the BH (or to the MBH, for MBH-bound debris); we then count the number of particles $N_{\Delta t}$ returning within $[t, t+\Delta t]$, so that $dM/dt = N_{\Delta t}\Delta m/\Delta t$. Because the debris has no self-gravity and by assumption exerts no force on the BHs, the two BHs follow their own hyperbolic two-body orbit, and all particles are advanced independently in a single integration. This
neglects the $\sim10\%$ backreaction of the debris ($\sim1\,M_\odot\sim0.1\,M_{\rm BH}$) on the BH, as well as the self-gravity of the stream. Hydrodynamical simulations show that the latter can be significant in $\mu$TDEs, leading to stream fragmentation~\citep{2021MNRAS.503.6005W}.

In Fig.~\ref{fig:debris} we show a snapshot (see also the online animation) of the system at $t=+260.58\,\rm days$ after the $\mu$TDE. At this time, about 12\% of the debris is bound to the BH ($\varepsilon_{\rm BH}<0$), $\approx8\%$ is unbound from the BH but bound to the MBH ($\varepsilon_{\rm BH}>0$ and $\varepsilon_{\bullet}<0$), and $\approx80\%$ is unbound from both BHs ($\varepsilon_{\rm BH}>0$ and $\varepsilon_\bullet>0$). The $\mu$TDE streams return and form a mini-disk around the BH within a few hours, while the $\approx8\%$ bound to the MBH returns to it after $\approx50\,\rm days$ (time of peak fallback). The debris falls back onto the MBH faster than the frozen-in approximation predicts, owing to the time-dependent tidal pull of the receding BH~\citep{2019ApJ...883L..17C,2021MNRAS.503.6005W}. The peak mass return rate onto the BH is $\approx0.5\,M_\odot\,\rm day^{-1}$, about three orders of magnitude higher than the peak mass fallback rate onto the MBH.

\subsection{Double flares and time delay}
\label{sec:time-delay}

The $\mu$TDE feeds two accretors: the $10\,M_\odot$ BH companion (prompt, fast mini-disk fallback) and the MBH (delayed, slower return). The total fallback curve therefore has two peaks, separated by the time delay $\Delta t_{\rm delay}$, which sets the characteristic timescale between the two flares. The time delay is dominated by the time it takes for the debris to return to the pericenter of the perturber and is estimated by $\Delta t_{\rm delay}\approx2\pi G M_\bullet (2|\varepsilon|_{\rm edge})^{-3/2}$, where $|\varepsilon|_{\rm edge}$ is the 90th percentile of the frozen-in energy distribution of the MBH-bound debris. The frozen-in approximation allows us to avoid tracking $10^4$ debris particles numerically in all $\approx9\times10^5$ $\mu$TDE simulations. For the benchmark case of Fig.~\ref{fig:debris}, the frozen-in time delay ($\approx120\,\rm days$) exceeds that of the direct ballistic integration ($\approx50\,\rm days$) by a factor of $\approx2.5$, so the delays reported here should be regarded as upper limits by a comparable factor.

\begin{figure}
    \centering
    \includegraphics[width=\linewidth]{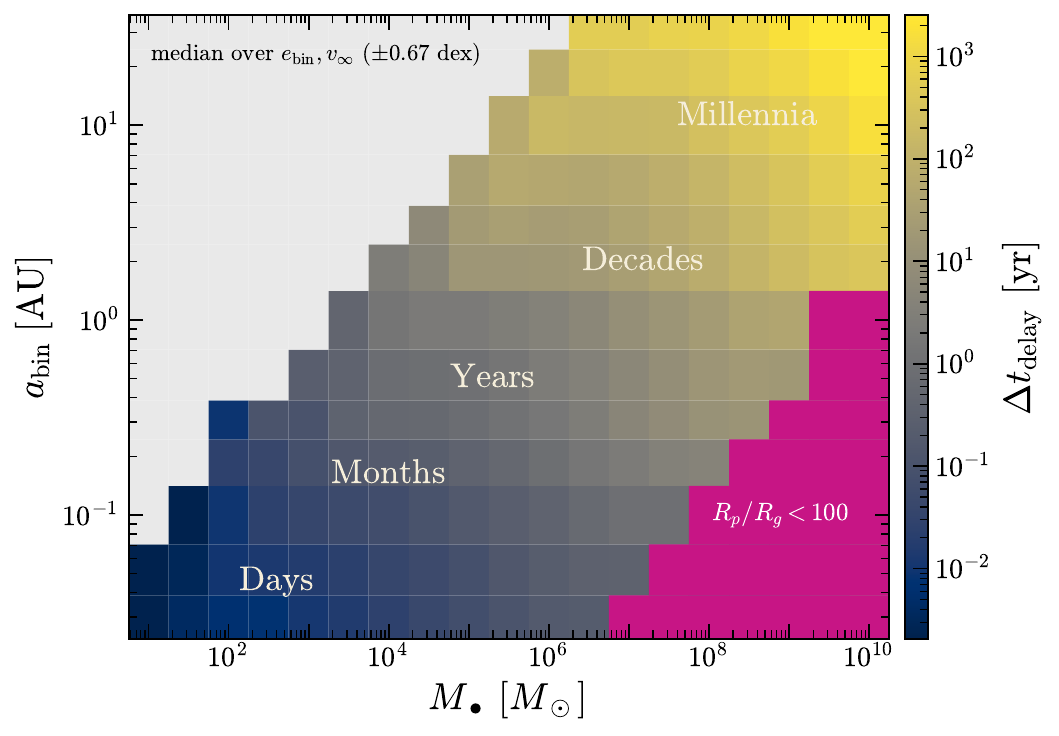}
    \caption{Heat map of the time delay $\Delta t_{\rm delay}$ between the $\mu$TDE and the peak fallback onto the perturber in the $a_{\rm bin}$--$M_\bullet$ parameter space assuming the frozen-in approximation. Marginalization over $e_{\rm bin}$ and $v_\infty$ results in a median scatter of $\pm0.67\,\rm dex$ about the values reported in the colorbar. There are no $\mu$TDE outcomes in the top left gray region. The lower right magenta region corresponds to relativistic encounters for which the binary's center of mass comes within $100\,R_g$ and our Newtonian integration scheme for the debris becomes inadequate.}
    \label{fig:time-delay}
\end{figure}

We perform a random forest feature importance test~\citep{Breiman2001} as implemented in \href{https://scikit-learn.org/stable/}{\tt scikit-learn}~\citep{Pedregosa2011} for the dependence of $\Delta t_{\rm delay}$ on $M_\bullet$, $a_{\rm bin}$, $e_{\rm bin}$, and $v_\infty$. Our results allow us to reduce the complexity of $\Delta t_{\rm delay}$ to only as a function of $M_\bullet$ ($\approx12\%$) and $a_{\rm bin}$ (at $\approx80\%$), while the dependence on $e_{\rm bin}$ and $v_\infty$ is weaker at $\approx8\%$ and $<1\%$, respectively. Marginalization over these nuanced parameters results in a scatter of $\pm0.67\,\rm dex$ in the $a_{\rm bin}$--$M_\bullet$ parameter space (see Fig.~\ref{fig:time-delay}). We exclude the relativistic regime (the magenta region in Fig.~\ref{fig:time-delay}) for which $R_p<100\,R_g$ and where the Newtonian approximation becomes inadequate.

Since the accretion happens onto BHs with, potentially, vastly different masses, we expect correspondingly different energetics in the two flare peaks: a faster rise and decay 
for the first $\mu$TDE event \citep{2021MNRAS.503.6005W, Ryu2022, Kremer2023}, which peaks in the X-rays/UV and may be accompanied by a gamma-ray burst~\citep{Perets2016}, and a slower, delayed transient with a longer-timescale decay peaking in the optical/UV and powered by accretion onto the MBH. Such a double-flare event should look like a fast luminous flare followed by a second, more classical-TDE-like flare at the same location on the sky.

The time delay is expected to range from days to months (months to years) for intermediate-mass BHs $M_\bullet=10^2$--$10^5\,M_\odot$ ($M_\bullet=10^5$--$10^7\,M_\odot$).
As an order-of-magnitude estimate, in the weakly penetrating regime, the fallback timescale can be obtained from Kepler's third law and using the fact that the energy spread in the debris is $\Delta \varepsilon = v_p\Delta v$, where $v_p=(2GM_\bullet/R_p)^{1/2}$ is the velocity at the pericenter and $\Delta v\approx v_{\rm orb}$ is the velocity spread in the debris; this yields, after some algebra,
\begin{align}
    \Delta t_{\rm delay} &\approx \frac{P_{\rm orb}(a_{\rm bin})}{(8\beta)^{3/4}}\left(\frac{M_\bullet}{M_{\rm bin}}\right)^{1/2}\nonumber\\
    &\simeq \frac{7.2\,\mathrm{months}}{\beta^{3/4}}\left(\frac{a_{\rm bin}}{0.1\,\mathrm{AU}}\right)^{3/2}\left(\frac{M_\bullet}{10^6\,M_\odot}\right)^{1/2},
    \label{eq:Dt-delay-eq}
\end{align}
for $\beta\lesssim1.26$, where the kinetic term $v_p\Delta v$ in $\Delta\varepsilon$ dominates over the potential term $GM_\bullet a_{\rm bin}/R_p^2$. Since most $\mu$TDEs occur for $\beta\lesssim1$ (Fig.~\ref{fig:branching_all}), this approximation is largely adequate.
For $\beta=0.6$ this estimate gives $\Delta t_{\rm delay}\approx10.5\,\rm months$, and for $M_\bullet=10^3\,M_\odot$, it is $\approx10\,\rm days$.
We provide an empirical fit to $\Delta t_{\rm delay}(M_\bullet, a_{\rm bin})$, under the frozen-in approximation, in Appendix~\ref{app:time-delay}, which can be used to infer $M_\bullet$ and $a_{\rm bin}$ from a doubly peaked light curve with characteristics as described above. We leave hydrodynamical simulations of the debris fallback to future work.

\subsection{Nuclear $\mu$TDE rates}

\begin{figure}
    \centering
    \includegraphics[width=\linewidth]{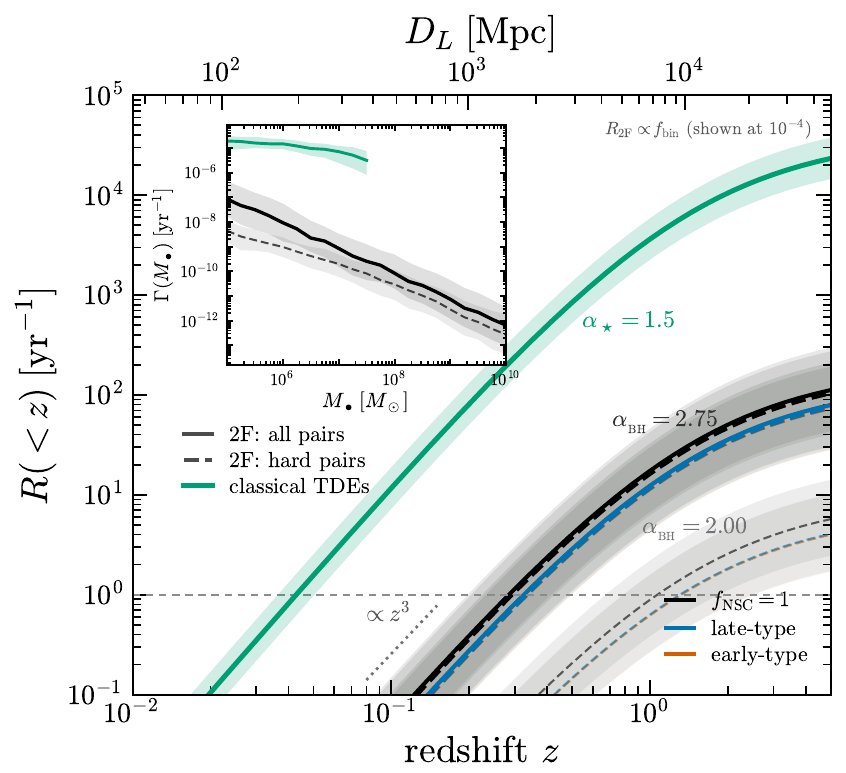}
    \caption{Cumulative observer-frame intrinsic rate of classical TDEs (green), with stellar density cusp index $\alpha_\star=1.5$, and double flares (2F) for $\alpha_{\rm BH}=2.75$ (strong segregation) and $\alpha_{\rm BH}=2.00$ (shallower segregation), as labeled. Solid (dashed) 2F curves include all (only hard) BH--star pairs. Black, blue, and orange curves assume an NSC occupation fraction $f_{\rm NSC}=1$, and the late-type and early-type $f_{\rm NSC}(M_{\rm gal})$ of~\citet{2020A&ARv..28....4N}, respectively. The 2F rate is proportional to the fraction of BH--star binaries, $f_{\rm bin}$, which is normalized to $10^{-4}$ here. The shaded bands for the 2F (classical TDE) predictions are dominated by the NSC mass (BH mass function normalization) scatter at $0.65\,\rm dex$ ($0.20\,\rm dex$). The inset shows the source-frame loss cone rates per galaxy. The classical TDE rate per galaxy stops at $M_\bullet\approx4\times10^7\,M_\odot$, while such a cutoff does not apply for 2F events.}
    \label{fig:rates}
\end{figure}

We estimate $\mu$TDE and classical TDE rates from BH--star--MBH and single-star--MBH encounters, respectively, using the loss cone theory of~\citet{1999MNRAS.306...35S}. 

Given the cross section $\Sigma$ for a particular process that removes objects from the surrounding NSC, the angular size of the loss cone at radius $r$ is $\theta_{\rm lc}^2\approx\Sigma/(\pi r^2)$. For $\mu$TDEs from BH--star--MBH encounters we use the cross section fitted in Appendix~\ref{app:muTDE-cross-section}, while for classical TDEs we use $\Sigma=2\pi G M_\bullet^{4/3}R_\star/(M_\star^{1/3}v_\infty^2)$~\citep{Stone:2016ryd}. We average the $\mu$TDE cross section over the statistical distributions $p(a_{\rm bin},e_{\rm bin})$ of~\citet{2019Natur.576..406S} for BH--star binaries, assuming the thermal, low-angular-momentum, and high-angular-momentum regimes. We further consider two cases: one in which we cap $a_{\rm bin}$ to retain only hard BH--star binaries, and another in which we include soft binaries.
We precompute $\Sigma(M_\bullet,v_\infty)$ on a grid for each regime and hard/all, and interpolate.

In the diffusive regime, the loss cone rate per object (or flux) is ${\cal F}_{\rm diff}\approx [\ln(2/\theta_{\rm lc})t_r]^{-1}$, where $t_r$ is the relaxation time at radius $r$ from the central MBH. In the full loss cone regime, the flux becomes ${\cal F}_{\rm full}\approx\theta_{\rm lc}^2/t_d$, where $t_d$ is the dynamical time at radius $r$.
We consider a harmonic interpolation between the diffusive and full loss cone regimes and write for the effective loss cone flux ${\cal F}(r)=[{\cal F}_{\rm diff}^{-1}(r) + {\cal F}_{\rm full}^{-1}(r)]^{-1}$. Finally, the total loss cone rate is $\Gamma_{\rm lc} = 4\pi\int_0^\infty {\cal F}(r)r^2n(r)\, dr$, where $n(r)$ is the number density of objects whose loss rate we wish to compute.

The total mass of the NSC, $M_{\rm NSC}$, surrounding the MBH is estimated by combining the $M_\bullet$--$M_{\rm gal}$~\citep[Fig.~3, left panel, ``All'']{2020ARA&A..58..257G} and $M_{\rm NSC}$--$M_{\rm gal}$~\citep[Eq.~1]{2020A&ARv..28....4N} relations, including both uncertainties and scatter, where $M_{\rm gal}$ is the galaxy stellar mass.

We follow~\citet{Alexander:2008tq} for the steady-state number density profile $n(r)\propto r^{-\alpha}$ of various species in the NSC. Stars relax to a power law shallower than Bahcall--Wolf, $\alpha_\star=1.5$ (but see~\citet{Rom:2026uma} for the effect of collisions in bringing this number down to 1.25), while heavier species, such as BH--star binaries, segregate toward the central MBH, with $2.0\le\alpha_{\rm BH}\le2.75$. If $N_\star=M_{\rm NSC}/M_\star$ is the total number of stars in the NSC, then $N_{\rm BH-\star}=N_\star f_{\rm bin}$, where $f_{\rm bin}$ is the fraction of stars that have a BH companion. Since $f_{\rm bin}$ is unconstrained, we use a conservative fiducial value of $f_{\rm bin}=10^{-4}$, and our results scale linearly with it.

We restrict our focus to MBHs with $M_\bullet>10^5\,M_\odot$, for which we have better observational constraints. We show the resulting double flare (2F) and classical TDE loss cone rates per galaxy in the inset of Fig.~\ref{fig:rates}. We find that $\Gamma_{2\rm F}\approx10^{-3}\Gamma_{\rm TDE}\approx10^{-8}\,\rm yr^{-1}$ at $M_\bullet=10^6\,M_\odot$. The 2F rate per galaxy can be fit by power laws $10^{-10.1\pm0.2}{\,\rm yr}^{-1}(M_{\bullet}/10^8\,M_\odot)^{-1.03\pm 0.12}$ and $10^{-10.7 \pm 0.2}{\,\rm yr}^{-1}(M_{\bullet}/10^8\,M_\odot)^{-0.8 \pm 0.1}$, for $\alpha_{\rm BH}=2.75$ and $\alpha_{\rm BH}=2.0$, with a scatter of $0.7\,\rm dex$ and $0.6\,\rm dex$, respectively.
The cumulative rates in Fig.~\ref{fig:rates} are insensitive to the choice of $p(a_{\rm bin},e_{\rm bin})$ regime, and we adopt the thermal one as fiducial.

Furthermore, assuming three models for the NSC occupation fraction $f_{\rm NSC}(M_\bullet)=f_{\rm NSC}(M_{\rm gal}(M_\bullet))$ from the right panel of Fig.~3 in~\citet{2020A&ARv..28....4N}, we convolve the rate per galaxy with the BH mass function from~\citet[Fig.~6, left panel]{2020ARA&A..58..257G} to compute the volumetric source-frame rate density of double flares and classical TDE flares. Finally, integrating over comoving volume and accounting for cosmological time dilation, under the assumption of a constant rate density, we obtain the cumulative number of double flares and classical TDE flares as a function of redshift (Fig.~\ref{fig:rates}).  We adopt a flat $\Lambda$CDM cosmology with the Planck 2018 parameters ($H_0=67.66~{\rm km\,s^{-1}\,Mpc^{-1}}$, $\Omega_{\rm m}=0.3097$;~\citealt{Planck2020}).
The nearest double flare each year could have a luminosity distance of a few $\rm Gpc$, and in the pessimistic scenario of shallow mass segregation of BH--star pairs the closest double flare could be at $\sim10\,\rm Gpc$. Under the strong (shallow) segregation assumption, we expect a total of $\sim100$ (a few) double flares per year within redshift $z\approx4$ (luminosity distance $\approx40\,\rm Gpc$), around half of which will have source-frame delay timescales under a year (a lower limit, given the frozen-in overestimate of Sec.~\ref{sec:time-delay}). The cusp slope of BH--star binaries and the binary fraction $f_{\rm bin}$ ($10^{-4}$ here) are the dominant sources of uncertainty.

\section{Summary and Discussion}
\label{sec:discussion}

We have investigated the dynamical production of transients from BH--star binaries undergoing hyperbolic encounters with MBHs in galactic nuclei. Unlike conventional TDEs, in which the MBH directly disrupts a star, here the MBH's tidal field perturbs the binary's internal orbit,
potentially driving it to high eccentricity and triggering a $\mu$TDE in which the star is disrupted by its BH companion. The resulting accretion first powers a flare near the stellar-mass BH, while the subsequent fallback of a fraction of the debris onto the MBH produces a second flare from the same sky location, delayed by weeks to months for $M_\bullet=10^{5}$--$10^{6}\,M_\odot$.

 Another outcome of the encounter is an exchange in which the BH is captured into a compact orbit around the MBH, producing an extreme mass-ratio inspiral (EMRI) and ejecting a hypervelocity star. Since the ejection velocity scales with $M_{\rm bin}^{-1/6}M_{\rm cap}^{1/2}$, where $M_{\rm cap}$ is the mass of the captured body, it exceeds that produced by the classical star--star--MBH Hills mechanism by a factor of $\simeq2.4$ for the same binary separation and penetration parameter~\citep{Hills1988}.
Because BH exchanges and $\mu$TDEs occur with comparable frequencies, the corresponding cross sections are also similar, with the BH and the star equally likely to be exchanged for $M_\bullet\gtrsim100\,M_\odot$.
We therefore expect EMRI rates comparable to those of $\mu$TDEs. Of these inspirals, we find that approximately 10--20\% merge within a Hubble time (estimated from the post-capture orbits in our simulations, evolved under gravitational-wave emission; \citealt{Peters:1964zz}), making them potential sources for the upcoming Laser Interferometer Space Antenna (LISA).

We predict approximately one double flare for every few hundred classical TDEs. If these events occur at the rates predicted by our model, the Vera C. Rubin Observatory should be able to reveal this previously unexplored population. The observational identification of double flares would have significant scientific implications. It would provide a means of constraining both the fraction of BH--star binaries and their spatial distribution within NSCs. In addition, the second flare, produced by the fallback of debris onto the MBH, could exhibit characteristic signatures that uniquely identify $\mu$TDEs, which may otherwise be difficult to distinguish when occurring in isolation.

\begin{acknowledgments}
We thank Fangyuan Yu for discussions on the simulation setup of~\citet{YuLai2024}. KK thanks Mitchell Karmen for discussions on the rates and observations of TDEs.
KK also thanks Daniel Kasen, Wenbin Lu, and Siva Darbha for discussions on $\mu$TDE radiative processes and fallback mass rates.
Support for this work was provided by NASA through the NASA Hubble Fellowship Grant No.~HST-HF2-51608.001-A awarded by the Space Telescope Science Institute, which is operated by the Association of Universities for Research in Astronomy, Inc., for NASA, under contract NAS5-26555.
The authors thank Stony Brook Research Computing and Cyberinfrastructure, and the Institute for Advanced Computational Science at Stony Brook University for access to the SeaWulf computing system, made possible by grants from the National Science Foundation (\#1531492 and Major Research Instrumentation award \#2215987), with matching funds from Empire State Development's Division of Science, Technology and Innovation (NYSTAR) program (contract C210148).
This work made use of Anthropic's Claude models (Opus~4.8 and Opus~5.5; \citealt{claude_opus_4_8,claude_opus_5_5})
via Claude Code for assistance with code development, data visualization, and aspects of the
computational workflow, including the $\mu$TDE debris-fallback pipeline, figure generation,
running and post-processing simulations, and preparation of the public code repository.
All scientific results were verified by the authors, and any remaining errors are the
authors' own.
\end{acknowledgments}

\section*{Data Availability}
The three-body simulation code, the $\mu$TDE debris-fallback pipeline, and the scripts used
to generate the figures in this work are publicly available at
\url{https://github.com/Kkritos/smbh-binary}. The simulation outputs (including the sampled
encounter parameters) are archived on Zenodo \citep{Kritos2026Data}.

\defcitealias{2022ApJ...935..167A}{Astropy Collaboration et al.\ 2022}
\software{\href{https://github.com/YihanWangAstro/SpaceHub}{\texttt{SpaceHub}}~\citep{Wang2021}, {\tt scikit-learn}~\citep{Pedregosa2011}, {\tt NumPy}~\citep{harris2020array}, {\tt SciPy}~\citep{2020SciPy-NMeth}, {\tt Matplotlib}~\citep{Hunter:2007}, {\tt h5py}~\citep{andrew_collette_2023_7560547}, {\tt pandas}~\citep{reback2020pandas}, {\tt astropy}~\citepalias{2022ApJ...935..167A}.}

\bibliography{biblio}

@ARTICLE{Gao2026,
       author = {{Gao}, Hao and {Lei}, Hanlun},
        title = "{Eccentricity and Inclination Excitation of Stellar Orbits around merging Black-Hole Binaries}",
      journal = {arXiv e-prints},
         year = 2026,
        month = sep,
          eid = {arXiv:2609.25109},
        pages = {arXiv:2609.25109},
          doi = {10.48550/arXiv.2609.25109},
archivePrefix = {arXiv},
       eprint = {2609.25109},
 primaryClass = {astro-ph.HE},
       adsurl = {https://scixplorer.org/abs/2026arXiv260925109G}
}

@ARTICLE{2022ApJ...935..167A,
       author = {{Astropy Collaboration} and {Price-Whelan}, Adrian M. and {Lim}, Pey Lian and {Earl}, Nicholas and {Starkman}, Nathaniel and {Bradley}, Larry and {Shupe}, David L. and {Patil}, Aarya A. and {Corrales}, Lia and {Brasseur}, C.~E. and et al.},
        title = "{The Astropy Project: Sustaining and Growing a Community-oriented Open-source Project and the Latest Major Release (v5.0) of the Core Package}",
      journal = {\apj},
         year = 2022,
        month = aug,
       volume = {935},
       number = {2},
          eid = {167},
        pages = {167},
          doi = {10.3847/1538-4357/ac7c74},
archivePrefix = {arXiv},
       eprint = {2206.14220},
 primaryClass = {astro-ph.IM},
       adsurl = {https://scixplorer.org/abs/2022ApJ...935..167A}
}

@software{reback2020pandas,
    author       = {The pandas development team},
    title        = {pandas-dev/pandas: Pandas},
    month        = feb,
    year         = 2020,
    publisher    = {Zenodo},
    version      = {latest},
    doi          = {10.5281/zenodo.3509134},
    url          = {https://doi.org/10.5281/zenodo.3509134}
}

@software{andrew_collette_2023_7560547,
  author    = {Andrew Collette and Thomas Kluyver and Thomas A Caswell and
               James Tocknell and Jerome Kieffer and Aleksandar Jelenak and
               Anthony Scopatz and Darren Dale and Chen and Thomas VINCENT and
               Matt Einhorn and payno and juliagarriga and Pierlauro Sciarelli and
               Valentin Valls and Satrajit Ghosh and Ulrik Kofoed Pedersen and
               Mark Kittisopikul and jakirkham and Martin Raspaud and
               Cyril Danilevski and Hameer Abbasi and John Readey and
               Kai M{\"u}hlbauer and Andrey Paramonov and Lawrence Chan and
               Robin De Schepper and V. Armando Sol{\'e} and jialin and
               Daniel Hay Guest},
  title     = {h5py/h5py: 3.8.0},
  month     = jan,
  year      = 2023,
  publisher = {Zenodo},
  version   = {3.8.0},
  doi       = {10.5281/zenodo.7560547},
  url       = {https://doi.org/10.5281/zenodo.7560547}
}

@Article{Hunter:2007,
  Author    = {Hunter, J. D.},
  Title     = {Matplotlib: A 2D graphics environment},
  Journal   = {Computing in Science \& Engineering},
  Volume    = {9},
  Number    = {3},
  Pages     = {90--95},
  publisher = {IEEE COMPUTER SOC},
  doi       = {10.1109/MCSE.2007.55},
  year      = 2007
}

@Article{         harris2020array,
 title         = {Array programming with {NumPy}},
 author        = {Charles R. Harris and K. Jarrod Millman and St{\'{e}}fan J.
                 van der Walt and Ralf Gommers and Pauli Virtanen and David
                 Cournapeau and Eric Wieser and Julian Taylor and Sebastian
                 Berg and Nathaniel J. Smith and Robert Kern and Matti Picus
                 and Stephan Hoyer and Marten H. van Kerkwijk and Matthew
                 Brett and Allan Haldane and Jaime Fern{\'{a}}ndez del
                 R{\'{i}}o and Mark Wiebe and Pearu Peterson and Pierre
                 G{\'{e}}rard-Marchant and Kevin Sheppard and Tyler Reddy and
                 Warren Weckesser and Hameer Abbasi and Christoph Gohlke and
                 Travis E. Oliphant},
 year          = {2020},
 month         = sep,
 journal       = {Nature},
 volume        = {585},
 number        = {7825},
 pages         = {357--362},
 doi           = {10.1038/s41586-020-2649-2},
 publisher     = {Springer Science and Business Media {LLC}},
 url           = {https://doi.org/10.1038/s41586-020-2649-2}
}

@ARTICLE{2020SciPy-NMeth,
  author  = {Virtanen, Pauli and Gommers, Ralf and Oliphant, Travis E. and
            Haberland, Matt and Reddy, Tyler and Cournapeau, David and
            Burovski, Evgeni and Peterson, Pearu and Weckesser, Warren and
            Bright, Jonathan and {van der Walt}, St{\'e}fan J. and
            Brett, Matthew and Wilson, Joshua and Millman, K. Jarrod and
            Mayorov, Nikolay and Nelson, Andrew R. J. and Jones, Eric and
            Kern, Robert and Larson, Eric and Carey, C J and
            Polat, {\.I}lhan and Feng, Yu and Moore, Eric W. and
            {VanderPlas}, Jake and Laxalde, Denis and Perktold, Josef and
            Cimrman, Robert and Henriksen, Ian and Quintero, E. A. and
            Harris, Charles R. and Archibald, Anne M. and
            Ribeiro, Ant{\^o}nio H. and Pedregosa, Fabian and
            {van Mulbregt}, Paul and {SciPy 1.0 Contributors}},
  title   = {{{SciPy} 1.0: Fundamental Algorithms for Scientific
            Computing in Python}},
  journal = {Nature Methods},
  year    = {2020},
  volume  = {17},
  pages   = {261--272},
  adsurl  = {https://rdcu.be/b08Wh},
  doi     = {10.1038/s41592-019-0686-2},
}

@article{Peters:1964zz,
    author = "Peters, P. C.",
    title = "{Gravitational Radiation and the Motion of Two Point Masses}",
    doi = "10.1103/PhysRev.136.B1224",
    journal = "Phys. Rev.",
    volume = "136",
    pages = "B1224--B1232",
    year = "1964"
}

@ARTICLE{Maeda2026,
       author = {{Maeda}, Kei-ichi and {Okawa}, Hirotada},
        title = "{Dynamics of a binary system around a supermassive black hole: Binary scattering and eccentric von Zeipel-Lidov-Kozai oscillations}",
      journal = {\prd},
         year = 2026,
        month = oct,
       volume = {114},
       number = {8},
          eid = {084078},
        pages = {084078},
          doi = {10.1103/jddn-gbnz},
archivePrefix = {arXiv},
       eprint = {2605.22525},
 primaryClass = {gr-qc},
       adsurl = {https://scixplorer.org/abs/2026PhRvD.114h4078M}
}

@ARTICLE{WangH2026,
       author = {{Wang}, Hao and {Liu}, Bin and {Samsing}, Johan},
        title = "{Black Hole Binary Mergers in Galactic Nuclei: Combined Effects of Secular Dynamics and Stellar Flybys}",
      journal = {\apj},
         year = 2026,
        month = nov,
       volume = {1011},
       number = {2},
          eid = {205},
        pages = {205},
          doi = {10.3847/1538-4357/aeb499},
archivePrefix = {arXiv},
       eprint = {2610.19879},
 primaryClass = {astro-ph.HE},
       adsurl = {https://scixplorer.org/abs/2026ApJ..1011..205W}
}

@article{Planck2020,
  author  = {{Planck Collaboration} and {Aghanim}, N. and {Akrami}, Y. and {Ashdown}, M. and others},
  title   = {Planck 2018 results. VI. Cosmological parameters},
  journal = {A\&A},
  year    = {2020},
  volume  = {641},
  pages   = {A6},
  doi     = {10.1051/0004-6361/201833910},
  eprint  = {1807.06209}
}

@ARTICLE{Ryu2022,
       author = {{Ryu}, Taeho and {Perna}, Rosalba and {Wang}, Yi-Han},
        title = "{Close encounters of stars with stellar-mass black hole binaries}",
      journal = {\mnras},
         year = 2022,
        month = oct,
       volume = {516},
       number = {2},
        pages = {2204-2217},
          doi = {10.1093/mnras/stac2316},
archivePrefix = {arXiv},
       eprint = {2206.00603},
 primaryClass = {astro-ph.HE},
       adsurl = {https://scixplorer.org/abs/2022MNRAS.516.2204R}
}

@ARTICLE{Kremer2023,
       author = {{Kremer}, Kyle and {Mockler}, Brenna and {Piro}, Anthony L. and {Lombardi}, James C.},
        title = "{Wind-reprocessed transients from stellar-mass black hole Tidal Disruption Events}",
      journal = {\mnras},
         year = 2023,
        month = oct,
       volume = {524},
       number = {4},
        pages = {6358-6373},
          doi = {10.1093/mnras/stad2239},
archivePrefix = {arXiv},
       eprint = {2305.08905},
 primaryClass = {astro-ph.HE},
       adsurl = {https://scixplorer.org/abs/2023MNRAS.524.6358K}
}

@ARTICLE{Langis2026,
       author = {{Langis}, D.~A. and {Liodakis}, I. and {Koljonen}, K.~I.~I. and {Paggi}, A. and {Globus}, N. and {Wyrzykowski}, L. and {Miko{\l}ajczyk}, P.~J. and {Kotysz}, K. and {Zieli{\'n}ski}, P. and {Ihanec}, N. and et al.},
        title = "{Repeating flares, X-ray outbursts and delayed infrared emission: A comprehensive compilation of optical tidal disruption events: TDECat}",
      journal = {\aap},
         year = 2026,
        month = mar,
       volume = {707},
          eid = {A171},
        pages = {A171},
          doi = {10.1051/0004-6361/202554882},
archivePrefix = {arXiv},
       eprint = {2506.05476},
 primaryClass = {astro-ph.HE},
       adsurl = {https://scixplorer.org/abs/2026A&A...707A.171L}
}

@ARTICLE{Johnson2026,
       author = {{Johnson}, Ian P.~A. and {Ryu}, Taeho and {Perna}, Rosalba},
        title = "{Hydrodynamic simulations of tidal disruption encores}",
      journal = {\prd},
         year = 2026,
        month = feb,
       volume = {113},
       number = {4},
          eid = {043032},
        pages = {043032},
          doi = {10.1103/3g9b-967g},
archivePrefix = {arXiv},
       eprint = {2510.23729},
 primaryClass = {astro-ph.HE},
       adsurl = {https://ui.adsabs.harvard.edu/abs/2026PhRvD.113d3032J}
}

@misc{Kritos2026Data,
  author       = {{Kritos}, Konstantinos and {Perna}, Rosalba},
  title        = {{Simulation data for ``Tidally Induced Coda: Double Transients from Black
                   Hole--Star Binary Encounters with Massive Black Holes''}},
  year         = 2026,
  month        = nov,
  publisher    = {Zenodo},
  version      = {1.0.0},
  doi          = {10.5281/zenodo.24372716},
  url          = {https://doi.org/10.5281/zenodo.24372716},
  note         = {Data set}
}

@ARTICLE{2019MEEP....7....1I,
       author = {{Ito}, Takashi and {Ohtsuka}, Katsuhito},
        title = "{The Lidov-Kozai Oscillation and Hugo von Zeipel}",
      journal = {Monographs on Environment, Earth and Planets},
         year = 2019,
        month = nov,
       volume = {7},
       number = {1},
        pages = {1-113},
          doi = {10.5047/meep.2019.00701.0001},
archivePrefix = {arXiv},
       eprint = {1911.03984},
 primaryClass = {astro-ph.EP},
       adsurl = {https://ui.adsabs.harvard.edu/abs/2019MEEP....7....1I}
}

@ARTICLE{2019ApJ...883L..17C,
       author = {{Coughlin}, Eric R. and {Nixon}, C.~J.},
        title = "{Partial Stellar Disruption by a Supermassive Black Hole: Is the Light Curve Really Proportional to t $^{-9/4}$?}",
      journal = {\apjl},
         year = 2019,
        month = sep,
       volume = {883},
       number = {1},
          eid = {L17},
        pages = {L17},
          doi = {10.3847/2041-8213/ab412d},
archivePrefix = {arXiv},
       eprint = {1907.03034},
 primaryClass = {astro-ph.GA},
       adsurl = {https://ui.adsabs.harvard.edu/abs/2019ApJ...883L..17C}
}

@misc{claude_opus_4_8,
  author       = {{Anthropic}},
  title        = {{Claude Opus 4.8}},
  year         = {2026},
  url          = {https://www.anthropic.com/claude},
  note         = {Large language model, accessed 2026}
}

@misc{claude_opus_5_5,
  author       = {{Anthropic}},
  title        = {{Claude Opus 5.5}},
  year         = {2026},
  url          = {https://www.anthropic.com/claude},
  note         = {Large language model, accessed November 2026}
}

@ARTICLE{2020A&ARv..28....4N,
       author = {{Neumayer}, Nadine and {Seth}, Anil and {B{\"o}ker}, Torsten},
        title = "{Nuclear star clusters}",
      journal = {\aapr},
         year = 2020,
        month = jul,
       volume = {28},
       number = {1},
          eid = {4},
        pages = {4},
          doi = {10.1007/s00159-020-00125-0},
archivePrefix = {arXiv},
       eprint = {2001.03626},
 primaryClass = {astro-ph.GA},
       adsurl = {https://ui.adsabs.harvard.edu/abs/2020A&ARv..28....4N}
}

@ARTICLE{2020ARA&A..58..257G,
       author = {{Greene}, Jenny E. and {Strader}, Jay and {Ho}, Luis C.},
        title = "{Intermediate-Mass Black Holes}",
      journal = {\araa},
         year = 2020,
        month = aug,
       volume = {58},
        pages = {257-312},
          doi = {10.1146/annurev-astro-032620-021835},
archivePrefix = {arXiv},
       eprint = {1911.09678},
 primaryClass = {astro-ph.GA},
       adsurl = {https://ui.adsabs.harvard.edu/abs/2020ARA&A..58..257G}
}

@article{Stone:2016ryd,
    author = {Stone, Nicholas C. and K{\"u}pper, Andreas H. W. and Ostriker, Jeremiah P.},
    title = "{Formation of Massive Black Holes in Galactic Nuclei: Runaway Tidal Encounters}",
    eprint = "1606.01909",
    archivePrefix = "arXiv",
    primaryClass = "astro-ph.GA",
    doi = "10.1093/mnras/stx097",
    journal = "Mon. Not. Roy. Astron. Soc.",
    volume = "467",
    number = "4",
    pages = "4180--4199",
    year = "2017"
}

@ARTICLE{2019Natur.576..406S,
       author = {{Stone}, Nicholas C. and {Leigh}, Nathan W.~C.},
        title = "{A statistical solution to the chaotic, non-hierarchical three-body problem}",
      journal = {\nat},
         year = 2019,
        month = dec,
       volume = {576},
       number = {7787},
        pages = {406-410},
          doi = {10.1038/s41586-019-1833-8},
archivePrefix = {arXiv},
       eprint = {1909.05272},
 primaryClass = {astro-ph.GA},
       adsurl = {https://ui.adsabs.harvard.edu/abs/2019Natur.576..406S}
}

@article{Rom:2026uma,
    author = "Rom, Barak and Sari, Re'em",
    title = "{Dynamics in Nuclear Stellar Clusters: The Impact of Collisions and Disrupted Binaries}",
    eprint = "2607.13152",
    archivePrefix = "arXiv",
    primaryClass = "astro-ph.GA",
    doi = "10.3847/2041-8213/ae7076",
    journal = "Astrophys. J. Lett.",
    volume = "1007",
    number = "1",
    pages = "L2",
    year = "2026"
}

@article{Alexander:2008tq,
    author = "Alexander, Tal and Hopman, Clovis",
    title = "{Strong mass segregation around a massive black hole}",
    eprint = "0808.3150",
    archivePrefix = "arXiv",
    primaryClass = "astro-ph",
    doi = "10.1088/0004-637X/697/2/1861",
    journal = "Astrophys. J.",
    volume = "697",
    pages = "1861--1869",
    year = "2009"
}

@ARTICLE{2021MNRAS.503.6005W,
       author = {{Wang}, Yi-Han and {Perna}, Rosalba and {Armitage}, Philip J.},
        title = "{Partial tidal disruption events by stellar mass black holes: Gravitational instability of stream and impact from remnant core}",
      journal = {\mnras},
         year = 2021,
        month = jun,
       volume = {503},
       number = {4},
        pages = {6005-6015},
          doi = {10.1093/mnras/stab802},
archivePrefix = {arXiv},
       eprint = {2103.09238},
 primaryClass = {astro-ph.HE},
       adsurl = {https://ui.adsabs.harvard.edu/abs/2021MNRAS.503.6005W}
}

@ARTICLE{1989ApJ...346L..13E,
       author = {{Evans}, Charles R. and {Kochanek}, Christopher S.},
        title = "{The Tidal Disruption of a Star by a Massive Black Hole}",
      journal = {\apjl},
         year = 1989,
        month = nov,
       volume = {346},
        pages = {L13},
          doi = {10.1086/185567},
       adsurl = {https://ui.adsabs.harvard.edu/abs/1989ApJ...346L..13E}
}

@article{Breiman2001,
  author  = {Breiman, Leo},
  title   = {Random Forests},
  journal = {Machine Learning},
  volume  = {45},
  number  = {1},
  pages   = {5--32},
  year    = {2001},
  doi     = {10.1023/A:1010933404324}
}

@article{Pedregosa2011,
  author  = {Pedregosa, F. and Varoquaux, G. and Gramfort, A. and Michel, V.
             and Thirion, B. and Grisel, O. and Blondel, M. and Prettenhofer, P.
             and Weiss, R. and Dubourg, V. and Vanderplas, J. and Passos, A.
             and Cournapeau, D. and Brucher, M. and Perrot, M. and Duchesnay, E.},
  title   = {Scikit-learn: Machine Learning in {P}ython},
  journal = {Journal of Machine Learning Research},
  volume  = {12},
  pages   = {2825--2830},
  year    = {2011}
}

@article{Karmen:2026ixp,
    author = "Karmen, Mitchell and Gezari, Suvi and Norman, Colin and Guolo, Muryel",
    title = "{Tidal Disruption Event Rates across Cosmic Time: Forecasts for LSST, Roman, and JWST and Their Constraints on the Supermassive Black Hole Mass Function}",
    eprint = "2602.04947",
    archivePrefix = "arXiv",
    primaryClass = "astro-ph.HE",
    doi = "10.3847/1538-4357/ae7a49",
    journal = "Astrophys. J.",
    volume = "1006",
    number = "1",
    pages = "20",
    year = "2026"
}

@article{Gezari:2021bmb,
    author = "Gezari, Suvi",
    title = "{Tidal Disruption Events}",
    eprint = "2104.14580",
    archivePrefix = "arXiv",
    primaryClass = "astro-ph.HE",
    doi = "10.1146/annurev-astro-111720-030029",
    journal = "Ann. Rev. Astron. Astrophys.",
    volume = "59",
    pages = "21--58",
    year = "2021"
}

@ARTICLE{Ryu2024,
       author = {{Ryu}, Taeho and {Perna}, Rosalba and {Cantiello}, Matteo},
        title = "{Tidal Disruption Encores}",
      journal = {\apjl},
         year = 2024,
        month = apr,
       volume = {965},
       number = {2},
          eid = {L25},
        pages = {L25},
          doi = {10.3847/2041-8213/ad3946},
archivePrefix = {arXiv},
       eprint = {2402.15590},
 primaryClass = {astro-ph.HE},
       adsurl = {https://ui.adsabs.harvard.edu/abs/2024ApJ...965L..25R}
}

@ARTICLE{Kozai1962,
       author = {{Kozai}, Yoshihide},
        title = "{Secular perturbations of asteroids with high inclination and eccentricity}",
      journal = {\aj},
         year = 1962,
        month = nov,
       volume = {67},
        pages = {591-598},
          doi = {10.1086/108790},
       adsurl = {https://ui.adsabs.harvard.edu/abs/1962AJ.....67..591K}
}

@ARTICLE{Fragione2020,
       author = {{Fragione}, Giacomo and {Metzger}, Brian D. and {Perna}, Rosalba and {Leigh}, Nathan W.~C. and {Kocsis}, Bence},
        title = "{Electromagnetic transients and gravitational waves from white dwarf disruptions by stellar black holes in triple systems}",
      journal = {\mnras},
         year = 2020,
        month = jun,
       volume = {495},
       number = {1},
        pages = {1061-1072},
          doi = {10.1093/mnras/staa1192},
archivePrefix = {arXiv},
       eprint = {1908.00987},
 primaryClass = {astro-ph.HE},
       adsurl = {https://ui.adsabs.harvard.edu/abs/2020MNRAS.495.1061F}
}

@ARTICLE{Fragione2019,
       author = {{Fragione}, Giacomo and {Leigh}, Nathan W.~C. and {Perna}, Rosalba and {Kocsis}, Bence},
        title = "{Tidal disruption events on to stellar black holes in triples}",
      journal = {\mnras},
         year = 2019,
        month = oct,
       volume = {489},
       number = {1},
        pages = {727-737},
          doi = {10.1093/mnras/stz2213},
archivePrefix = {arXiv},
       eprint = {1905.09471},
 primaryClass = {astro-ph.GA},
       adsurl = {https://ui.adsabs.harvard.edu/abs/2019MNRAS.489..727F}
}

@ARTICLE{Lidov1962,
       author = {{Lidov}, M.~L.},
        title = "{The evolution of orbits of artificial satellites of planets under the action of gravitational perturbations of external bodies}",
      journal = {\planss},
         year = 1962,
        month = oct,
       volume = {9},
       number = {10},
        pages = {719-759},
          doi = {10.1016/0032-0633(62)90129-0},
       adsurl = {https://ui.adsabs.harvard.edu/abs/1962P&SS....9..719L}
}

@ARTICLE{Hills1988,
   author = {{Hills}, J.~G.},
    title = "{Hyper-velocity and tidal stars from binaries disrupted by a massive Galactic black hole}",
  journal = {Nature},
     year = 1988,
   volume = {331},
    pages = {687-689},
      doi = {10.1038/331687a0}
}

@ARTICLE{Rees1988,
   author = {{Rees}, M.~J.},
    title = "{Tidal disruption of stars by black holes of 10 to the 6th-10 to the 8th solar masses in nearby galaxies}",
  journal = {Nature},
     year = 1988,
   volume = {333},
    pages = {523-528},
      doi = {10.1038/333523a0}
}

@ARTICLE{Ginsburg2007,
   author = {{Ginsburg}, I. and {Loeb}, A.},
    title = "{Hypervelocity collisions of binary stars at the Galactic Centre}",
  journal = {\mnras},
     year = 2007,
   volume = {376},
    pages = {492-497},
      doi = {10.1111/j.1365-2966.2007.11444.x}
}

@ARTICLE{Antonini2010,
   author = {{Antonini}, F. and {Faber}, J. and {Gualandris}, A. and {Merritt}, D.},
    title = "{Tidal Breakup of Binary Stars at the Galactic Center and Its Consequences}",
  journal = {\apj},
     year = 2010,
   volume = {713},
    pages = {90-104},
      doi = {10.1088/0004-637X/713/1/90}
}

@ARTICLE{Antonini2011,
   author = {{Antonini}, F. and {Lombardi}, Jr., J.~C. and {Merritt}, D.},
    title = "{Tidal Encounters of Binary and Single Stars in Galactic Nuclei}",
  journal = {\apj},
     year = 2011,
   volume = {731},
    pages = {128},
      doi = {10.1088/0004-637X/731/2/128}
}

@ARTICLE{Bradnick2017,
   author = {{Bradnick}, B. and {Mandel}, I. and {Levin}, Y.},
    title = "{Stellar binaries in galactic nuclei: tidally stimulated mergers followed by tidal disruptions}",
  journal = {\mnras},
     year = 2017,
   volume = {469},
    pages = {2042-2048},
      doi = {10.1093/mnras/stx1007}
}

@ARTICLE{YuLai2024,
   author = {{Yu}, F. and {Lai}, D.},
    title = "{Binary Stars Approaching Supermassive Black Holes: Tidal Breakup, Double Stellar Disruptions, and Stellar Collision}",
  journal = {\apj},
     year = 2024,
   volume = {977},
    pages = {268},
      doi = {10.3847/1538-4357/ad93a6}
}

@ARTICLE{AntoniniPerets2012,
   author = {{Antonini}, F. and {Perets}, H.~B.},
    title = "{Secular Evolution of Compact Binaries near Massive Black Holes: Gravitational Wave Sources and Other Exotica}",
  journal = {\apj},
     year = 2012,
   volume = {757},
    pages = {27},
      doi = {10.1088/0004-637X/757/1/27}
}

@ARTICLE{Perets2016,
   author = {{Perets}, H.~B. and {Li}, Z. and {Lombardi}, J.~C. and {Milcarek}, S.~R.},
    title = "{Micro-tidal Disruption Events by Stellar Compact Objects and the Production of Ultra-long GRBs}",
  journal = {\apj},
     year = 2016,
   volume = {823},
    pages = {113},
      doi = {10.3847/0004-637X/823/2/113}
}

@ARTICLE{Wang2021,
   author = {{Wang}, Y.-H. and {Leigh}, N. and {Liu}, B. and {Perna}, R.},
    title = "{SpaceHub: A high-performance gravity integration toolkit for few-body problems in astrophysics}",
  journal = {\mnras},
     year = 2021,
   volume = {505},
    pages = {1053-1064},
      doi = {10.1093/mnras/stab1287}
}

@ARTICLE{Wang2021b,
       author = {{Wang}, Yi-Han and {McKernan}, Barry and {Ford}, Saavik and {Perna}, Rosalba and {Leigh}, Nathan W.~C. and {Mac Low}, Mordecai-Mark},
        title = "{Symmetry Breaking in Dynamical Encounters in the Disks of Active Galactic Nuclei}",
      journal = {\apjl},
         year = 2021,
        month = dec,
       volume = {923},
       number = {2},
          eid = {L23},
        pages = {L23},
          doi = {10.3847/2041-8213/ac400a},
archivePrefix = {arXiv},
       eprint = {2110.03698},
 primaryClass = {astro-ph.HE},
       adsurl = {https://ui.adsabs.harvard.edu/abs/2021ApJ...923L..23W}
}

@ARTICLE{1999MNRAS.306...35S,
       author = {{Syer}, D. and {Ulmer}, A.},
        title = "{Tidal disruption rates of stars in observed galaxies}",
      journal = {\mnras},
         year = 1999,
        month = jun,
       volume = {306},
       number = {1},
        pages = {35-42},
          doi = {10.1046/j.1365-8711.1999.02445.x},
archivePrefix = {arXiv},
       eprint = {astro-ph/9812389},
 primaryClass = {astro-ph},
       adsurl = {https://ui.adsabs.harvard.edu/abs/1999MNRAS.306...35S}
}

\appendix

\section{Collision fraction of star--star binary--MBH encounters}
\label{app:starstar-collisions}

\begin{figure}
    \centering
    \includegraphics[width=0.49\linewidth]{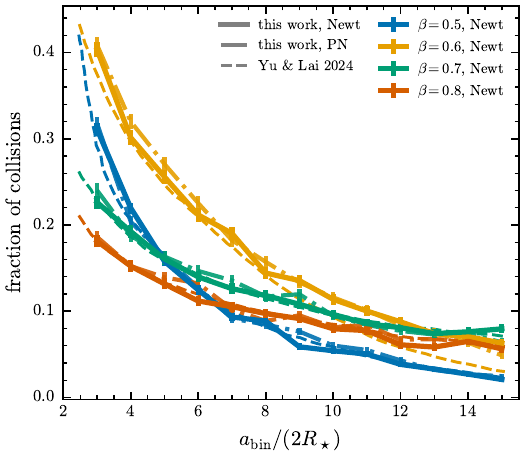}
    \includegraphics[width=0.49\linewidth]{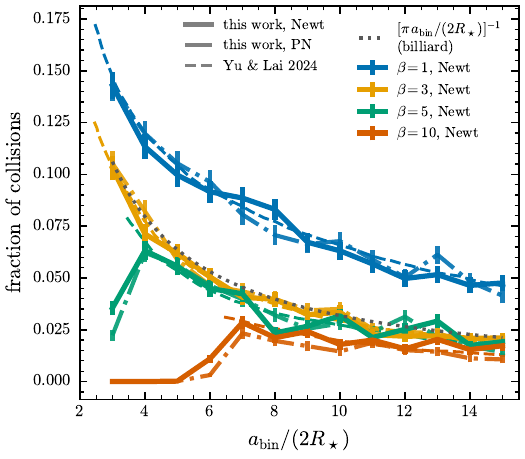}
    \caption{The fraction of star--star--MBH encounters that end up in a stellar collision for shallow ($\beta<1$, left) and deep ($\beta\ge1$, right) encounters. Our results are shown as thick solid lines (Newtonian only, ``Newt'') and thick dot-dashed lines (including 1\,PN, 2\,PN, and 2.5\,PN force terms, ``PN''), with binomial error bars. The thin dashed lines correspond to the results of~\citet{YuLai2024}. Finally, the thin dotted gray line in the right panel corresponds to the billiard-ball approximation of~\citet{YuLai2024}.}
    \label{fig:collision_fraction}
\end{figure}

In this Appendix, we study the fraction of collision outcomes in star--star--MBH encounters, as a function of the star--star binary's semimajor axis $a_{\rm bin}$ and the penetration parameter $\beta$. 
The possible outcomes in these encounters include flybys, collisions, exchanges, and bound triples. We fix $M_\bullet=10^6\,M_\odot$, $M_\star=1\,M_\odot$, $R_\star=1\,R_\odot$, and consider circular binaries as in Figs.~5 and~7 of~\citet{YuLai2024}. Moreover, in order to reproduce their collision fraction, we adopt the same sampling and initial conditions: $\beta\in\{0.5, 0.6, 0.7, 0.8, 1.0, 3.0, 5.0, 10.0\}$, $\alpha\equiv a_{\rm bin}/(2R_\star)\in\{3,\dots,15\}$ (integer steps), and inclination between the inner and outer orbits $\iota\in\{30^\circ,60^\circ,90^\circ\}$ (private communication with F. Yu).

We carry out 1000 realizations per $(\beta, \alpha,\iota)$ point by randomly drawing the phase of the binary and the longitude of the ascending node in $[0,2\pi)$, giving us an effective 3000 realizations per $(\beta,\alpha)$ configuration. Assuming isotropic orientations, the three inclinations are weighted by $\sin\iota$.
While~\citet{YuLai2024} considered only Newtonian forces, we repeat the same simulations including 1\,PN, 2\,PN, and 2.5\,PN terms for comparison.

We show the resulting collision fractions in Fig.~\ref{fig:collision_fraction}, where we distinguish between shallow ($\beta<1$, left panel) and deep ($\beta\ge1$, right panel) encounters, as in Figs.~5 and~7 of~\citet{YuLai2024}, respectively.
Our results agree with those of~\citet{YuLai2024} within the binomial errors. We also find that the PN terms change the collision fraction by less than 2\% in star--star--MBH encounters.
Nevertheless, we include the PN terms in our BH--star--MBH simulations, since they become important for $M_\bullet\gtrsim10^8\,M_\odot$, where they increase the fraction of direct relativistic plunges and captures.

We further note that for deep encounters ($\beta\ge1$) the collision fractions of~\citet{YuLai2024} cannot be extrapolated below the smallest $\alpha$ they consider. For increasing $\beta$, there is a threshold value for $\alpha$ below which the collision fraction drops to zero, because such deep encounters always lead to an exchange and subsequent ejection: the tidal field is strong enough to separate the stars promptly, leaving a vanishingly small loss cone for direct collision near pericenter.

\section{Fitting formulae for $\mu$TDE outcomes}

\subsection{Time delay between the first and second flares}
\label{app:time-delay}

\begin{figure}
    \centering
    \includegraphics[width=0.49\linewidth]{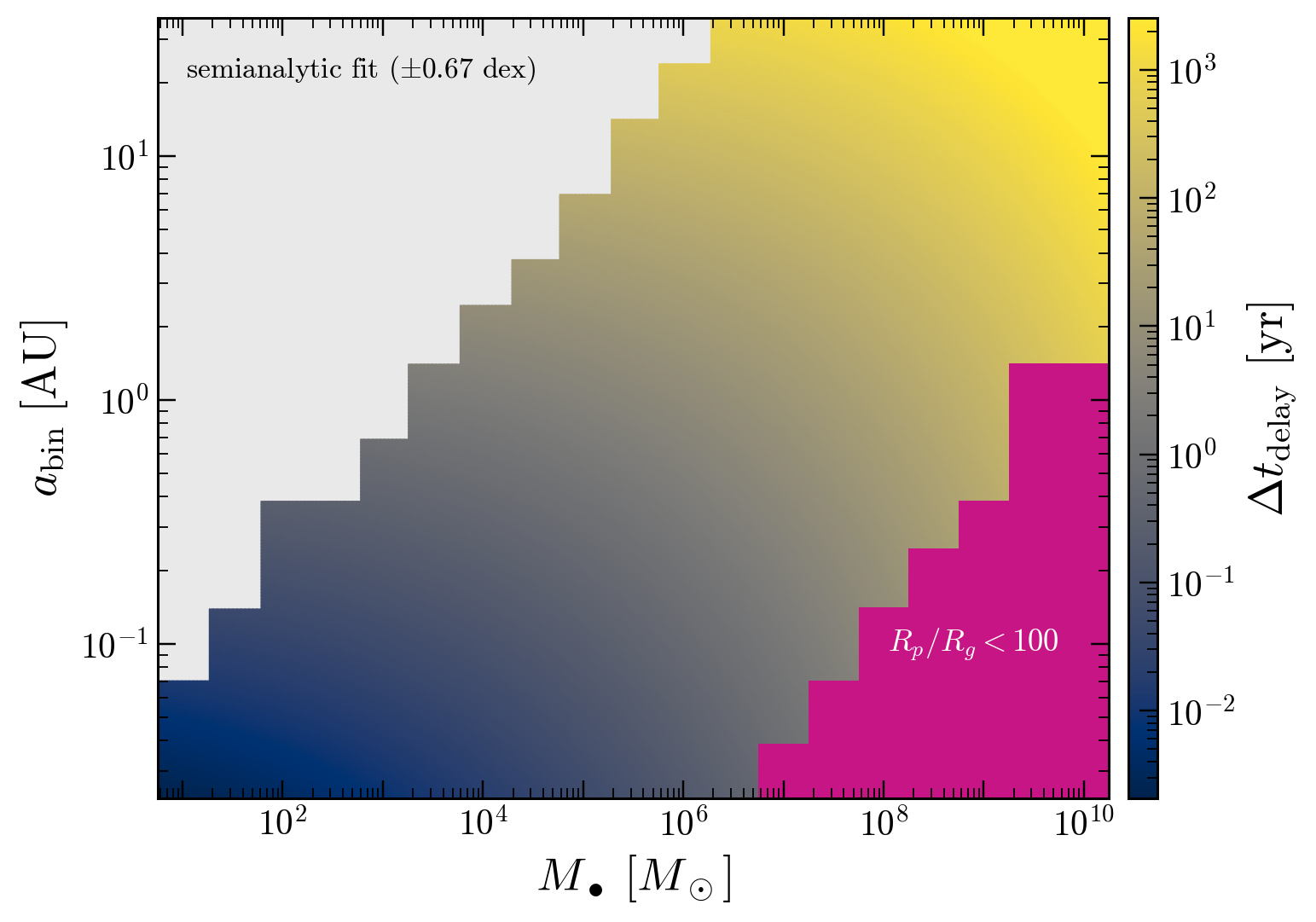}
    \includegraphics[width=0.49\linewidth]{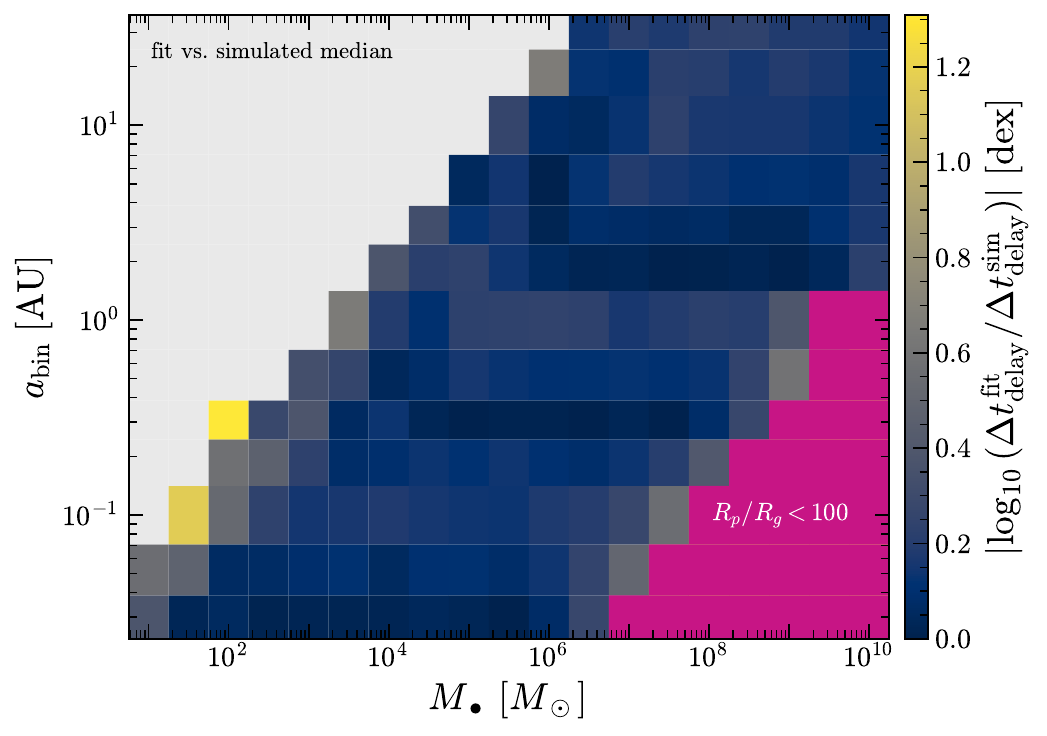}
    \caption{Empirical fit of the time delay, Eq.~\eqref{eq:Dt-fit} (left), and the residuals relative to the median shown in Fig.~\ref{fig:time-delay} (right). The fit excludes the magenta region where $R_p<100\,R_g$.}
    \label{fig:delay_fit_vs_numerical}
\end{figure}

We fit an empirical quadratic formula for the time delay between the two flares in a $\mu$TDE outcome near an MBH (computed under the frozen-in approximation; see Sec.~\ref{sec:time-delay}):
\begin{equation}
\log_{10}\left(\frac{\Delta t_{\rm delay}}{\rm yr}\right) = 0.2511-0.1181X_M + 2.0696X_a + 0.0353X_M^2 + 0.1251X_a^2 - 0.1473X_MX_a,
\label{eq:Dt-fit}
\end{equation}
where $X_M\equiv\log_{10}(M_\bullet/M_\odot)$ and $X_a\equiv\log_{10}(a_{\rm bin}/\rm AU)$.
The fit has a coefficient of determination $R^2=0.804$ and a scatter of $\pm0.67\,\rm dex$.
Including $e_{\rm bin}$ reduces the residual variance by only 3.7\%, so we retain only $M_\bullet$ and $a_{\rm bin}$ in the fit.
The empirical fit and residuals are shown in Fig.~\ref{fig:delay_fit_vs_numerical}.
The fit is valid only for $R_p\ge100\,R_g$.

\subsection{Fitting the cross section}
\label{app:muTDE-cross-section}

We fit the $\mu$TDE fraction $f_{\mu\rm TDE}$ by noting that a $\mu$TDE happens when the penetration parameter $\beta$ exceeds a critical value $\beta_{\rm crit}$. Since we sample $b^2$ uniformly, $f_{\mu\rm TDE}$ is the fraction of encounters with $\beta > \beta_{\rm crit}$, and is the ratio of the areas:
\begin{equation}
f_{\mu\rm TDE} = \frac{R_{\rm crit}^{\,2} + R_{\rm foc}\,R_{\rm crit}}{R_{p,\max}^{\,2} + R_{\rm foc}\,R_{p,\max}},
\label{eq:fmu-fit}
\end{equation}
where $R_{\rm crit}=R_{t,\rm bin}/\beta_{\rm crit}$, $R_{\rm foc}=2G(M_\bullet+M_{\rm bin})/v_\infty^2$, and
\begin{equation}
\log_{10}\beta_{\rm crit} = -0.379 + 1.567\,L - 0.239\,L^2 + 0.519\,e_{\rm bin} - 1.124\,e_{\rm bin}^2,\qquad
L \equiv \log_{10}\frac{a_{\rm bin}(1-e_{\rm bin})}{0.01\,\rm AU}.
\end{equation}
The fit has a coefficient of determination $R^2=0.921$ and is valid only for $R_{\rm crit}\le R_{p,\max}$. Neither $M_\bullet$ nor $v_\infty$ enters $\beta_{\rm crit}$ (adding them improves the fit by less than $0.3\%$); they enter only through $R_{t,\rm bin}$ and the focusing term. In the hard-binary limit, the cross section is dominated by the focusing term, and the $\mu$TDE fraction reduces to $f_{\mu\rm TDE}\approx(5\beta_{\rm crit})^{-1}$. 

\begin{figure}
    \centering
    \includegraphics[width=0.49\linewidth]{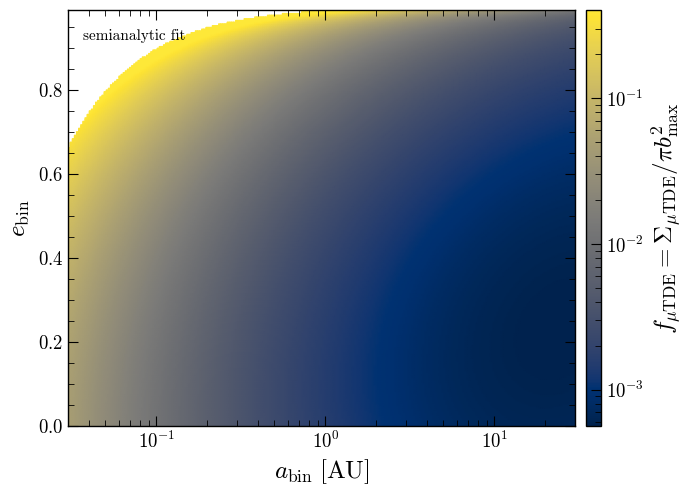}
    \includegraphics[width=0.49\linewidth]{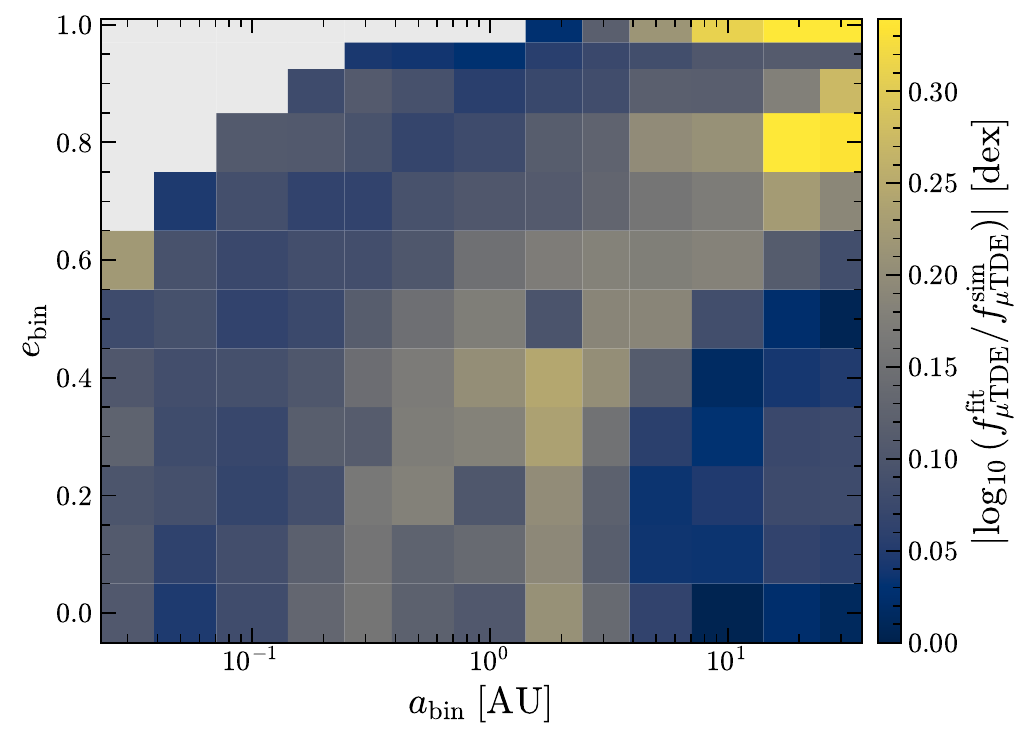}
    \caption{Semianalytic fit of Eq.~\eqref{eq:fmu-fit} for the fraction of $\mu$TDE outcomes $f_{\mu\rm TDE}$ in the $(a_{\rm bin}, e_{\rm bin})$ plane (left) and residuals relative to the simulations (right).}
    \label{fig:fmu_fit_vs_numerical}
\end{figure}

The fit reproduces the simulated $f_{\mu\rm TDE}$ (Fig.~\ref{fig:fmu_fit_vs_numerical}) with a median absolute deviation of 0.11 dex (85\% of grid cells within a factor of two, without systematic bias or residual dependence on $M_\bullet$ or $v_\infty$); it underestimates $f_{\mu\rm TDE}$ by up to a factor $\sim1.7$ for $M_\bullet\gtrsim10^{8.5}\,M_\odot$ and for $e_{\rm bin}\to0.99$.

\end{document}